\documentclass[twocolumn,nodate,showpacs,preprintnumbers,amsmath,amssymb,aps,prl]{revtex4-1}

\begin{document}

\title{Intrinsic time gravity and the Lichnerowicz-York equation }

\author{Niall \'O Murchadha}\email{niall@ucc.ie}

\affiliation{Physics Department, University College, Cork, Ireland}

\author{Chopin Soo}\email{cpsoo@mail.ncku.edu.tw}

\affiliation{Department of Physics, National Cheng Kung University, Taiwan}

\author{Hoi-Lai Yu}\email{hlyu@phys.sinica.edu.tw}

\affiliation{Institute of Physics, Academia Sinica, Taiwan}

\begin{abstract}

We investigate the effect on the Hamiltonian structure of general relativity of choosing an intrinsic time to fix the time slicing.  3-covariance with momentum constraint is maintained, but the Hamiltonian constraint is replaced by a dynamical equation for the trace of the momentum. This reveals a very simple structure with a local reduced Hamiltonian.  The theory is easily generalized; in particular, the square of the Cotton-York tensor density can be added as an extra part of the potential
while at the same time maintaining the classic 2 + 2 degrees of freedom. Initial data construction is simple in the extended theory; we get a generalized  Lichnerowicz-York equation with nice existence and uniqueness properties. Adding standard matter fields is quite straightforward.

\end{abstract}



\maketitle

\section{Introduction}

The Hamiltonian theory for  general relativity was most clearly laid out by Arnowitt, Deser, and Misner \cite{ADM} more than 50 years ago. The phase space consists of a pair
$(g_{ij}, \pi^{ij})$, where $g_{ij}$ is a Riemannian 3-metric and $\pi^{ij}$ is the conjugate momentum. These cannot be freely chosen because  they must satisfy the constraints
\begin{equation}\label{H}
-gR + \pi^{ij}\pi_{ij} - \frac{1}{2}\pi^2 = 0; \ \ \  {\rm and} \ \ \ \nabla_i\pi^{ij} = 0,
\end{equation}
where $\pi = g_{ij}\pi^{ij}$ is the trace of $\pi^{ij}$. These conditions are known respectively as the Hamiltonian and momentum constraints. We must also choose  a scalar and a vector
$(N, N^i)$. These are the `lapse' and `shift' . The total Hamiltonian, in the compact without boundary case, is
\begin{equation}\label{HH}
\textstyle{H=\int\big[g^{-1/2}N(-gR+\pi^{ij}\pi_{ij}-\frac{1}{2}\pi^2)-2N_j\nabla_i\pi^{ij}\big]d^3x}.
\end{equation}
Therefore $(N, N^i)$ are the Lagrange multipliers of the constraints. The first of Hamilton's equations gives the relationship between $\pi^{ij}$ and the time derivative of $g_{ij}$
\begin{equation}\label{dg/dt}
\frac{\partial g_{ij}}{\partial t} = 2Ng^{-1/2}(\pi_{ij} - \frac{1}{2}g_{ij}\pi) + \nabla_i N_j +\nabla_j N_i.
\end{equation}

A major difficulty with the canonical quantization of gravity program is this freedom to choose $N$\cite{Sooyu}. Each choice of $N$ gives a different slicing of spacetime, and
reflects the 4-covariance of the Einstein equations.  We want to break this covariance, choose a natural time variable, and compute the emergent lapse. James York in
\cite{York1} pointed out that the local volume, $\sqrt{g}$, and $\pi/\sqrt{g}$ are canonically conjugate, and suggested that $\pi/\sqrt{g}$ is a natural time, and that $\sqrt{g}$ is the
local energy. This is called an `extrinsic' time because $\pi^{ij}$ is essentially the extrinsic curvature of the slice. We have
$g^{-1/2}(\pi_{ij} - \frac{1}{2}g_{ij}\pi) = K_{ij}, $
where $K_{ij}$ is the extrinsic curvature. With York's choice, $\sqrt{g}$ is the associated energy density and the total volume of the slice is the Hamiltonian. Two years earlier
Charles Misner \cite{Misner} made the opposite choice, albeit in a minisuperspace context. He picked the local volume as his time and $\pi$ is then the energy density. This is
where he introduced the `mixmaster universe'. The great advantage of Misner's choice is that a local reduced Hamiltonian appears. Recently, two of us \cite{Sooyu} presented a theory
of gravity passing from classical to quantum regimes with a paradigm shift from 4-covariance to 3-covariance. Its framework revealed the primacy of a local reduced Hamiltonian and
intrinsic time proportional to $\ln g^{1/3}$. The present article is an expansion and extension of part of that work.

\section{Intrinsic time gravity}

We start by choosing the intrinsic time function as $\ln g^{1/3}$. Therefore $\delta \ln g^{1/3}$, for any variation, is a scalar. Further, the canonical conjugate is $\pi$, with
coefficient of unity. We split the metric into unimodular and determinant parts via $\bar{g}_{ij} = g^{-1/3}g_{ij}$, and split $\pi^{ij}$ into tracefree and trace parts via
$\bar\pi^{ij} = g^{1/3}(\pi^{ij} - g^{ij}\pi/3)$. The symplectic 1-form becomes

\begin{equation}
\int\, {\pi}^{ij}\delta g_{ij} = \int {\bar\pi}^{ij}\delta{\bar g}_{ij} + \pi \delta \ln g^{1/3}.
\end{equation}
We see immediately that $({\bar g}_{ij}, {\bar \pi}^{ij})$ and $(\ln g^{1/3}, \pi)$ form conjugate pairs, which is clearly the generalization of  the $P_i\delta x^i +P_0\delta t$ one gets in the case of a simple particle.
With the choice of time as $\ln g^{1/3}$ (which varies from $-\infty$ to $+\infty$, instead of $0$ to $\infty$), we can see that $\pi$ is analogous to $P_0 = -E$.

We now substitute the decomposition of $(g_{ij},\pi^{ij})$ into the Hamiltonian constraint, Eq.(\ref{H}), to give
\begin{equation}\label{H1}
-gR + \bar{g}_{ik}\bar{g}_{jl}{\bar\pi}^{ij}{\bar\pi}^{kl} - \beta^2 \pi^2 = 0.
\end{equation}
where $\beta^2 =  \frac{1}{6}$ for GR. It is a free positive parameter for the extended theory which we discuss later. We know that the Hamiltonian as the generator of time translation, is conjugate to time and should equal the energy,  and therefore the true Hamiltonian density with this choice of intrinsic time should be $-\pi$. We then just solve the Hamiltonian constraint, which is equivalent to $(\pi - \bar{H}/\sqrt{\beta^2}) (\pi + \bar{H}/\sqrt{\beta^2})=0$, to find the
reduced Hamiltonian
\begin{equation}\label{RH}
-\pi = \bar{H}/\beta  = \frac{1}{\beta}\sqrt{\bar{g}_{ik}\bar{g}_{jl}{\bar\pi}^{ij}{\bar\pi}^{kl} - gR}; \quad \beta =\pm\sqrt{\beta^2}.
\end{equation}

A simple toy model for this process is given by the relativistic particle, which satisfies a constraint
\begin{equation}
-(P^0)^2 + \vec{P}\cdot\vec{P} + m^2 = 0.
\end{equation}
$P^0$ is the energy, and therefore the physical Hamiltonian is
\begin{equation}
E = P^0 = -P_0 =  H  = \sqrt{ \vec{P}\cdot\vec{P} + m^2}.
\end{equation}
Hamilton's equations give the equations of motion for $\vec{P}$. We then add the constraint $P^0 = H$ as a dynamical equation to determine $P^0$.

The Hamiltonian ${\bar H}/\beta$ of  Eq.(\ref{RH}) generates $\ln g^{1/3}(x,t)$ translations, and gives equations of motion for $\bar{g}_{ij}$ and ${\bar\pi}^{ij}$ with respect to this intrinsic time variable. We need to write $gR$ in terms of $\bar{g}_{ij}$ and $\ln g^{1/3}$. This is
quite straightforward. We do not have an equation for $\ln g$, since it is the time, but we  do need a dynamical equation for $\pi$  which is given by
\begin{equation}\label{pi}
\pi + \frac{1}{\beta} \bar{H}(\bar{g}_{ij}, \bar{\pi}^{ij}, \ln g^{1/3}) =0.
\end{equation}
This is a rewriting of the Hamiltonian constraint, but it is no longer to be viewed as a constraint; rather, it is the evolution equation for $\pi$. This is the fundamental equation for intrinsic time gravity; and it can be interpreted respectively as the Hamilton-Jacobi equation and Schr\"odinger equation in the semi-classical and quantum regimes\cite{Sooyu}.

We choose the positive root for $\bar H$ when we take the square-root. Negative values of $\pi$  correspond, in line with current observations, to an expanding universe.
The reduced Hamiltonian can be constructed before solving for $\pi$. Since the system is 3-covariant, the evolution equations will propagate the
momentum constraint. This is the only constraint left since the Hamiltonian constraint is gone because we have a unique choice of (intrinsic) time. Thus, we are left with a system that has the expected
2 + 2 degrees of freedom.

The symplectic potential of the conjugate pair $(\ln g^{1/3}, \pi)$ contributes  $\int \int (\pi \frac{\partial \ln g^{1/3}}{\partial t})d^3x \delta t = -\int [\int \frac{\bar H}{\beta}\frac{\partial \ln g^{1/3}}{\partial t}d^3x ]dt$, to the action; and it is thus clear that $\int \frac{\bar H}{\beta}\frac{\partial \ln g^{1/3}}{\partial t}d^3x$ contributes to the total Hamiltonian generating $t$-translations, where $t$ is the ADM time parameter.
By subtracting the tangential change generated by the the momentum constraint, the rate of change of the normal component of $\ln g^{1/3}$ is
\begin{equation}
f= \lim_{\delta t \rightarrow 0} \frac{\delta \ln g^{1/3} - \pounds_{\vec{N}\delta t}\ln g^{1/3}}{  \delta t}= \frac{\partial \ln g^{1/3}}{  \partial t} - \frac{2}{3}\nabla_i N^i.
 \label{f}
\end{equation}
The classical evolution of $({\bar g}_{ij},{\bar \pi}^{ij})$ w.r.t. the ADM time variable $t$ can equivalently be obtained from the effective Hamiltonian
\begin{eqnarray}
&&H_f=\textstyle{\int \left[ f(\bar{H}/\beta)  +N^iH_i\right]d^3x}\nonumber\\
=&&\textstyle{\int\big[\frac{f}{\beta}\sqrt{\bar{g}_{ik}\bar{g}_{jl}{\bar\pi}^{ij}{\bar\pi}^{kl} - gR} - 2N_i(g^{-\frac{1}{3}}\nabla_j{\bar \pi}^{ij})\big]d^3x.}\label{RHf}
\end{eqnarray}

On  the other hand, by contracting Eq.(\ref{dg/dt}) with $g^{ij}$,  we find
\begin{equation}
\frac{\partial \ln g^{1/3}}{  \partial t}= - \frac{N\pi}{3\sqrt{g}} +\frac{2}{3}\nabla_i N^i . \label{N3}
\end{equation}
If we choose
\begin{equation}
N = \frac{-3f\sqrt{g}}{\pi} ,\label{N4}
\end{equation}
 it is clear that Eq.(\ref{f}) agrees with Eq.(\ref{N3}), and if one uses the lapse  as given by Eq.(\ref{N4}), it is a straightforward exercise to show that the evolution equations arising from $H_f$ agree with those from the ADM evolution equations.
 Given any $t$-foliation of spacetime, we can find a corresponding reduced Hamiltonian that generates it. This means that `many fingered time' lives on in the intrinsic time picture
 but in a very different form by the way the  emergent $N$, as given by Eq.(\ref{N4}), depends on the (freely chosen) $f$.

We have expressed GR in a very simple form:  we have only the effective Hamiltonian, Eq.(\ref{RHf}), combined with the dynamical equation for the trace of the momentum, Eq.(\ref{pi}), and the momentum constraint.

Moving away from vacuum GR, how can we generalize this structure, while maintaining the 2  + 2 degrees of freedom and the 3-covariance?

There are two changes we can make. We can multiply each of the three terms in the Hamiltonian constraint, as given by either Eq.(\ref{H}) or Eq.(\ref{H1}), with arbitrary constants. This means deforming $\beta$ in the $\pi^2$ term away from $\frac{1}{6}$, which is equivalent  to changing the constant in the DeWitt  supermetric\cite{DW}. We really only need two constants because we can always rescale one of them to unity. Thus we replace $gR$ by $\alpha^2gR$ and leave the $\bar{g}_{ik}\bar{g}_{jl}{\bar\pi}^{ij}{\bar\pi}^{kl}$ alone. More radically, we can replace the
`potential', $-R$, by any scalar function $V$ of the metric, and everything still works. The set of equations now reads
\begin{eqnarray}
H_f&=& \textstyle{\int\big(\frac{f}{\beta}\sqrt{\bar{g}_{ik}\bar{g}_{jl}{\bar\pi}^{ij}{\bar\pi}^{kl} + gV} +N^iH_i\big)d^3x}, \label{h2}\\
&&\pi  :=- \frac{1}{\beta}\sqrt{\bar{g}_{ik}\bar{g}_{jl}{\bar\pi}^{ij}{\bar\pi}^{kl} + gV},\label{pi2}\\
&& f := \frac{\partial g^{1/3}}{\partial t} - \frac{2}{3}\nabla_i N^i.\label{t2}
\end{eqnarray}

\section{Initial data and extensions of the Lichnerowicz-York equation}

It is not quite that easy, however; we need to find explicit initial data. There is no point in having a nice Hamiltonian system with no solutions. We need to find a pair
$({g}_{ij}, {\pi}^{ij})$ that satisfies both $\beta\pi + \bar{H} = 0$ and $\nabla_i\pi^{ij} = 0$. In the case of GR this pair reduces to the standard constraints; and the only really successful general way of solving them is the conformal method, which was initiated by Lichnerowicz \cite{Lich}. There is a very
comprehensive account in \cite{CB}, especially in Chapter VII. The technique is to choose free data that consist of a base metric $\hat{g}_{ij}$, a symmetric tensor density
$\hat{\pi}^{iTT}_j$ that is  both tracefree and divergence-free with respect to $\hat{g}$, and a scalar $\hat{p}$. It is particularly simple if $\hat{p}$ is a constant. This
guarantees that the extrinsic curvature has constant trace, and thus we construct a `constant mean curvature' (CMC) slice. We make a conformal transformation $g_{ij} =
\phi^4\hat{g}_{ij}, \pi^{ij} = \phi^{-4}\hat{\pi}^{ij}$.  It turns out that   $\hat{\pi}^{iTT}_j$ is conformally invariant; it remains $TT$ with respect to $g_{ij}$. Thus we write   ${\pi}^{iTT}_j =
\hat{\pi}^{iTT}_j$. We also set $\pi = \sqrt{g}\hat{p}$. We are guaranteed that $\pi^{ij} = (\pi^{ij}_{TT}  + \sqrt{g}g^{ij}\hat{p}/3)$ satisfies the momentum constraint for
any conformal factor $\phi$. In particular, we can specialize to $\phi^4=g^{1/3}$ to get $(\bar{g}_{ij}, \bar{\pi}^{ij})=(\hat{g}_{ij}, {\hat\pi}^{ij}_{TT})$.   We now
seek an appropriate $\phi$ in order to solve the Hamiltonian constraint. In GR, this reduces to solving the Lichnerowicz-York (LY) equation
\begin{equation}\label{LY}
8\hat{\nabla}^2\phi - \hat{R}\phi + \hat{g}^{-1}\hat{\pi}^{iTT}_j\hat{\pi}^{jTT}_i\phi^{-7} - \frac{1}{6}\hat{p}^2\phi^5 = 0,
\end{equation}
where $\hat{g}$ is the determinant of $\hat{g}_{ij}$. This is an extremely nice equation because it always has a unique, positive solution \cite{OMY}.

Let us now multiply the $R$ and the $\pi^2$ terms in the Hamiltonian constraint  by
  arbitrary positive constants. This means replacing Eq.(\ref{H1}) by
\begin{equation}
-\alpha^2gR + \bar{g}_{ik}\bar{g}_{jl}{\bar\pi}^{ij}{\bar\pi}^{kl} - \beta^2\pi^2 = 0.\label{H5}
\end{equation}
The new LY equation becomes
\begin{equation}\label{ly}
8\alpha^2\hat{\nabla}^2\phi - \alpha^2\hat{R}\phi + \hat{g}^{-1}\hat{\pi}^{iTT}_j\hat{\pi}^{jTT}_i\phi^{-7} - \beta^2\hat{p}^2\phi^5 = 0.
\end{equation}
This equation is just as nice as the original LY equation, Eq.(\ref{LY}), because it too always has a unique positive solution.
Such a rescaling  has been recently discussed in a different context in \cite{BOM}.

However, if we add any other function of the metric to Eq.(\ref{H5}) we destroy the nice properties of the LY equation. For example, if we were to add an $R^2$ term, the LY
equation would pick up the term $(8\nabla^2\phi - R\phi)^2$, which changes the nature of the LY equation completely and the existence and uniqueness results no longer hold.

There is one exception. York, in \cite{York2}, rediscovered a conformally covariant tensor which is a function of the metric. This is now known as the Cotton-York tensor(density), i.e.,
$\beta^{ij}$.
This transforms exactly as a TT tensor does under conformal transformations, so $\beta^i_j$ is conformally invariant. Therefore we can pick a generalized Hamiltonian as
\begin{equation}\label{H6}
-\alpha^2gR + \bar{g}_{ik}\bar{g}_{jl}{\bar\pi}^{ij}{\bar\pi}^{kl} +\gamma^2 \beta^i_j\beta^j_i- \beta^2\pi^2 = 0,
\end{equation}
where $\gamma$ is another coupling constant. The generalized LY equation becomes

\begin{eqnarray}\label{LY1}
8\alpha^2\hat{\nabla}^2\phi - \alpha^2\hat{R}\phi + \hat{g}^{-1}(&\hat{\pi}&^{ij}_{TT}\hat{\pi}^{TT}_{ij} + \gamma^2\hat{\beta}^i_j\hat{\beta}^j_i)\phi^{-7} \nonumber\\
 &-& \hat{g}^{-1}\beta^2\hat{p}^2\phi^5 = 0.
\end{eqnarray}
This is just as well-behaved as the original LY equation: it always possesses a positive unique solution. This conformal factor maps the free data onto a solution both of the
generalized Hamiltonian constraint, Eq.(\ref{H6}), and the momentum constraint.

We can do slightly better than this. Let us assume that we would like to have a more general Hamiltonian, and add an $R^2$ term to the potential. Thus, we would like initial data
which satisfies
\begin{equation}\label{H7}
-\alpha^2gR - \rho gR^2 + \bar{g}_{ik}\bar{g}_{jl}{\bar\pi}^{ij}{\bar\pi}^{kl}  +\gamma^2 \beta^i_j\beta^j_i- \beta^2\pi^2 = 0,
\end{equation}
where $\rho$ is another parameter.
The analogue of the LY equation will be
\begin{eqnarray}\label{LY2}
8\alpha^2\hat{\nabla}^2\phi &-&\alpha^2 \hat{R}\phi +  \hat{g}^{-1}(\hat{\pi}^{ij}_{TT}\hat{\pi}^{TT}_{ij} +\gamma^2\hat{\beta}^i_j\hat{\beta}^j_i)\phi^{-7} \nonumber -
\beta^2\hat{p}^2\phi^5\\ &=&  \rho \phi^{-5}(8\hat{\nabla}^2\phi -\hat{R}\phi)^2.
\end{eqnarray}
This, as mentioned earlier, is a deeply unpleasant equation. However, let us linearize it about $\rho = 0$. First, let us solve Eq.(\ref{LY1}) and use that conformal factor to map
the free data to a set that satisfies Eq.(\ref{H7}). This means that we are varying about $\phi_0 \equiv 1$. Then, differentiate Eq.(\ref{LY2}) with respect to $\rho$, and set $\rho = 0$ and
$\phi \equiv 1$. Denoting $d\phi/d\rho = h$, we get
\begin{equation}\label{LY3}
8\alpha^2\hat{\nabla}^2h - \alpha^2\hat{R}h -7\hat{g}^{-1}  (\hat{\pi}^{ij}_{TT}\hat{\pi}^{TT}_{ij} +\gamma^2\hat{\beta}^i_j\hat{\beta}^j_i)h - 5\beta^2\hat{p}^2h =
\hat{R}^2.\end{equation}
Note that the $h$ terms that should appear on the right hand side vanish because they are all multiplied by $\rho$.  Multiplying Eq.(\ref{H6}) by
$h/\hat{g}$ and subtracting from Eq.(\ref{LY3}) (to eliminate the term linear in $\hat{R}$) we get
\begin{equation}\label{LY4}
8\alpha^2\hat{\nabla}^2h -   (8\hat{g}^{-1}\hat{\pi}^{ij}_{TT}\hat{\pi}^{TT}_{ij} +8\gamma^2\hat{g}^{-1}\hat{\beta}^i_j\hat{\beta}^j_i + 4\beta^2p^2)h =
\hat{R}^2.\end{equation}
It is obvious that the coefficient of the undifferentiated $h$ in Eq.(\ref{LY4}) is negative. This means that the homogeneous equation has no kernel. In
turn, this means that the inhomogeneous equation, Eq.(\ref{LY4}), has a unique solution. This is the Fredholm alternative (see, for example \cite{MT}). Now we can use the
implicit function theorem (again, see \cite{MT}) to guarantee that the non-linear equation, Eq.(\ref{LY2}), has a solution for a range of $\rho$'s in a neighborhood of zero.
Unfortunately, this technique gives us no estimate as to the size of this neighborhood. One can immediately see that  this technique works for any choice of metric potential.

We construct CMC initial data, not because it plays any fundamental role in this theory, but simply out of convenience. The slices generated by the Hamiltonian, Eq.(\ref{h2}), will not stay
CMC. We can use the same technique, i.e., the Fredholm alternative + the implicit function theorem, to relax the condition that $\hat p$ is a constant. We can replace it by assuming
$\hat{p} = p_0 + \theta p_1$, where $p_0$ is a nonzero constant, $p_1$ is a function of position, and $\theta$ is a parameter. Now the conformal method gives us a system of 4 coupled nonlinear equations
\cite{OMY1}. If we linearize about $\theta = 0$, we find that the equations decouple and we get a system with the necessary existence and uniqueness properties. This means we can construct
non-CMC initial data, but it is not clear how large the deviation from CMC we can allow.

We can add a whole range of matter fields to the system. We can add a cosmological constant, a massive or massless scalar field, a Maxwell or a Yang-Mills field, dust, or
even neutrinos \cite{I}. The reader should note that the coupling constant of the Cotton-York density term in the generalized Hamiltonian, $\gamma^2\beta^i_j\beta^j_i$, is dimensionless in natural units (in Eq.(1) and elsewhere in this article we use units $2\kappa$=1).  In 3-covariant modifications of general relativity this term is essential to the perturbative power-counting renormalizability of the quantum theory\cite{Horava,Sooyu}.

At early (intrinsic) times, the $R$ term in Eq.(17) is suppressed by the $e^{ \ln g}$ factor, while the Cotton-York tensor density-squared term is conformally invariant and independent of  $g$; and at late times the theory becomes more and more like GR. Thus, near the Big Bang, with $\ln g \rightarrow -\infty$, the Cotton-York density term should dominate.  In particular, we expect that the BKL model \cite{BKL}, where the time variation of the gravitational field is expected to dominate over the spatial variations, will no longer hold in the presence of non-trivial amount of the Cotton-York tensor.

A key advantage the intrinsic time formalism has over the extrinsic time formalism is that the Hamiltonian constraint can be easily put in the form of an algebraic equation for $\pi$. If we wanted to use extrinsic time, we would need to solve the Hamiltonian constraint for $\sqrt{g}$.
Therefore the reduced Hamiltonian in the extrinsic time gauge is  a non-local object. Another sign of non-locality is that the lapse function for a CMC foliation is determined by
\begin{equation}
\nabla^2N - g^{-1}\pi^{ij}\pi_{ij}N = K(t).\label{N7}
\end{equation}
This is obviously a nice elliptic equation, but clearly non-local as distinct from   Eq.(\ref{N4}), the equation for the intrinsic time lapse.

It is clear that using an intrinsic time gives a very clean Hamiltonian structure in classical gravity. These good properties carry over when we implement the canonical quantization program\cite{Sooyu}.

\begin{acknowledgments}

This work has been supported in part by the National Science Council of Taiwan under Grant Nos. NSC101-2112-M-006-007-MY3, 97-2112-M-001-005-MY3, and the National Center
for Theoretical Sciences, Taiwan.

\end{acknowledgments}

\end{document}